\documentclass{pkas}

\def\year{2014} % publication year
\def\month{March} % publication month
\def\volume{00} % journal volume
\def\issue{0} % journal issue
\def\beginpage{1} % first page of article
\def\endpage{99} % last page of article
\def\received{---} % date paper was received by PKAS
\def\accepted{---} % date of acceptance
\date{Received \received ; accepted \accepted}

\title{
Akari, SCUBA2 and Herschel data of pre-stellar cores
}

\author[1]{D. Ward-Thompson} 
\author[1]{K. Pattle} 
\author[1]{J. M. Kirk}
\author[2]{P. Andr\'e}
\author[3]{J. Di Francesco}
\affil[1]{Jeremiah Horrocks Institute, University of Central Lancashire,
Preston, Lancashire, PR1 2HE, UK \email{dward-thompson@uclan.ac.uk,
kpattle@uclan.ac.uk, jmkirk@uclan.ac.uk}}
\affil[2]{Laboratoire d'Astrophysique, C.E. Saclay, Orme des Merisiers,
91191, Gif-sur-Yvette Cedex, France 
\email{philippe.andre@cea.fr}}
\affil[3]{Department of Physics and Astronomy, University of Victoria, 
P.O. Box 355, Victoria, BC, Canada
\email{james.difrancesco@nrc-cnrc.gc.ca}}
\def\corrauthor{
D. Ward-Thompson
}

\def\runningauthor{
Ward-Thompson et al.
}

\def\runningtitle{
Pre-stellar cores
}

\def\keywords{
stars: formation --- molecular clouds --- filaments --- cores: pre-stellar ---
protostars 
}

\def\abstracttext{
We show Akari data, Herschel data and data from the SCUBA2 camera 
on JCMT, of molecular clouds. We focus on pre-stellar cores within the clouds.
We present Akari data of the L1147-1157 ring in Cepheus and show how the
data indicate that the cores are being externally heated. We present SCUBA2
and Herschel data of the Ophiuchus region and show how the environment is
also affecting core evolution in this region.
We discuss the effects of the magnetic field in the Lupus I region, and 
how this lends support to a model for the formation and evolution of cores
in filamentary molecular clouds.
}

\begin{document}
\pkashead %% set title, authors, abstract, etc.

%%%%%%%%%%%%%%%%%%%%%%%%%%%%%%%%%%%%%%%%%%%%%%%%%%%%%%%%%%%%%%%%%%%%%
%%% BEGIN MAIN TEXT HERE %%%%%%%%%%%%%%%%%%%%%%%%%%%%%%%%%%%%%%%%%%%%
%%%%%%%%%%%%%%%%%%%%%%%%%%%%%%%%%%%%%%%%%%%%%%%%%%%%%%%%%%%%%%%%%%%%%

\section{Introduction}

Stars form in dense cores in molecular clouds.
Exactly how these cores 
form is still a matter of debate (e.g. \citealt{andre2014}). 
The cores which are gravitationally 
bound are known as pre-stellar cores (\citealt{wardthompson2007};
\citealt{difrancesco2007}), which then collapse to form Class 0 
protostars (\citealt{andre1993}).  
The mass function of cores can be modelled onto the IMF of stars
(\citealt{goodwin2008}). Hence cores are a significant stage in star 
formation. Recent work with the Akari
satellite, the Herschel Space Observatory and the SCUBA2 camera on the 
JCMT has led to new insight into core formation
(\citealt{andre2014}).

\section{Cepheus}

The molecular cloud in Cepheus has been studied by many authors (e.g. \citealt{kirk2009}). Figure~1
shows an extinction map of the region (\citealt{dobashi2005}). 
North is at the top, east is to the left.
The square towards the western edge of the image
shows the region we have studied. This is the L1147-L1157 ring.

Figure 2 shows the extinction map (thick contours)
of a close-up of the ring (\citealt{dobashi2005}), superposed on a greyscale of the Digitised Sky Survey. The
various molecular cloud cores from L1147 to L1157 are marked.
Also shown in thin contours on Figure 2 is the Akari 90-micron
emission from the region overlaid (\citealt{nutter2009}). The Akari emission
can be seen to be offset from the extinction peaks of both L1148 and L1155.
In fact the far-infrared emission seems to be wrapping itself around the 
dense material (traced by extinction). 

%%% FIGURE %%%%%%%%%%%%%%%%%%%%%%%%%%%%%%%%%%%%%%%%%%%%%%%%%%%%%%%%%%%%%%%%%%%%
\begin{figure}
\centering
\includegraphics[width=60mm]{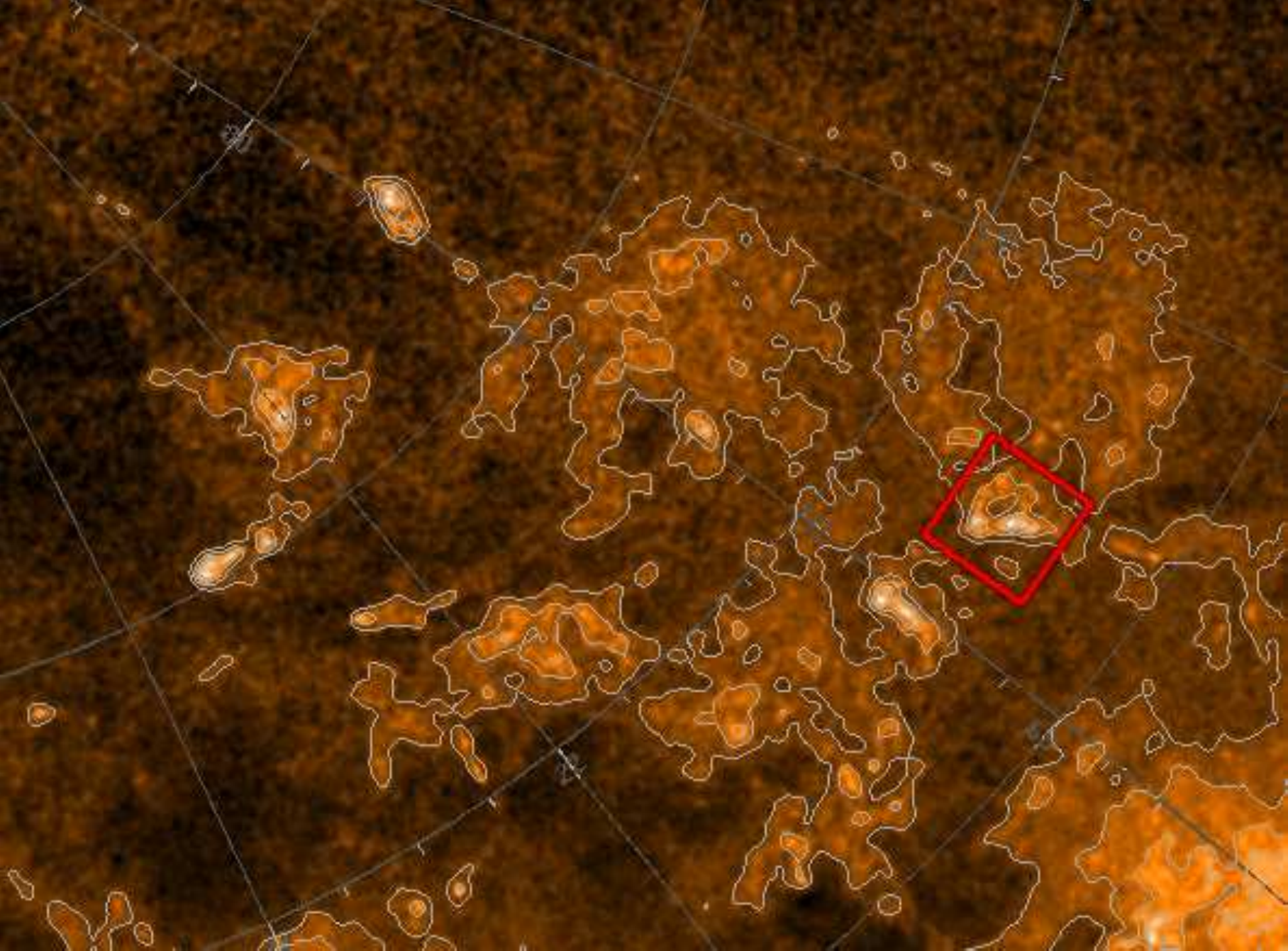}
\caption{An extinction map of the Cepheus region (from \citealt{dobashi2005}).
North is at the top, east is to the left.
The ring of cores being studied here is indicated by a box towards the western
edge of this image.}
\end{figure}
%%%%%%%%%%%%%%%%%%%%%%%%%%%%%%%%%%%%%%%%%%%%%%%%%%%%%%%%%%%%%%%%%%%%%%%%%%%%%%%

%%% FIGURE %%%%%%%%%%%%%%%%%%%%%%%%%%%%%%%%%%%%%%%%%%%%%%%%%%%%%%%%%%%%%%%%%%%%
\begin{figure}
\centering
\includegraphics[width=60mm]{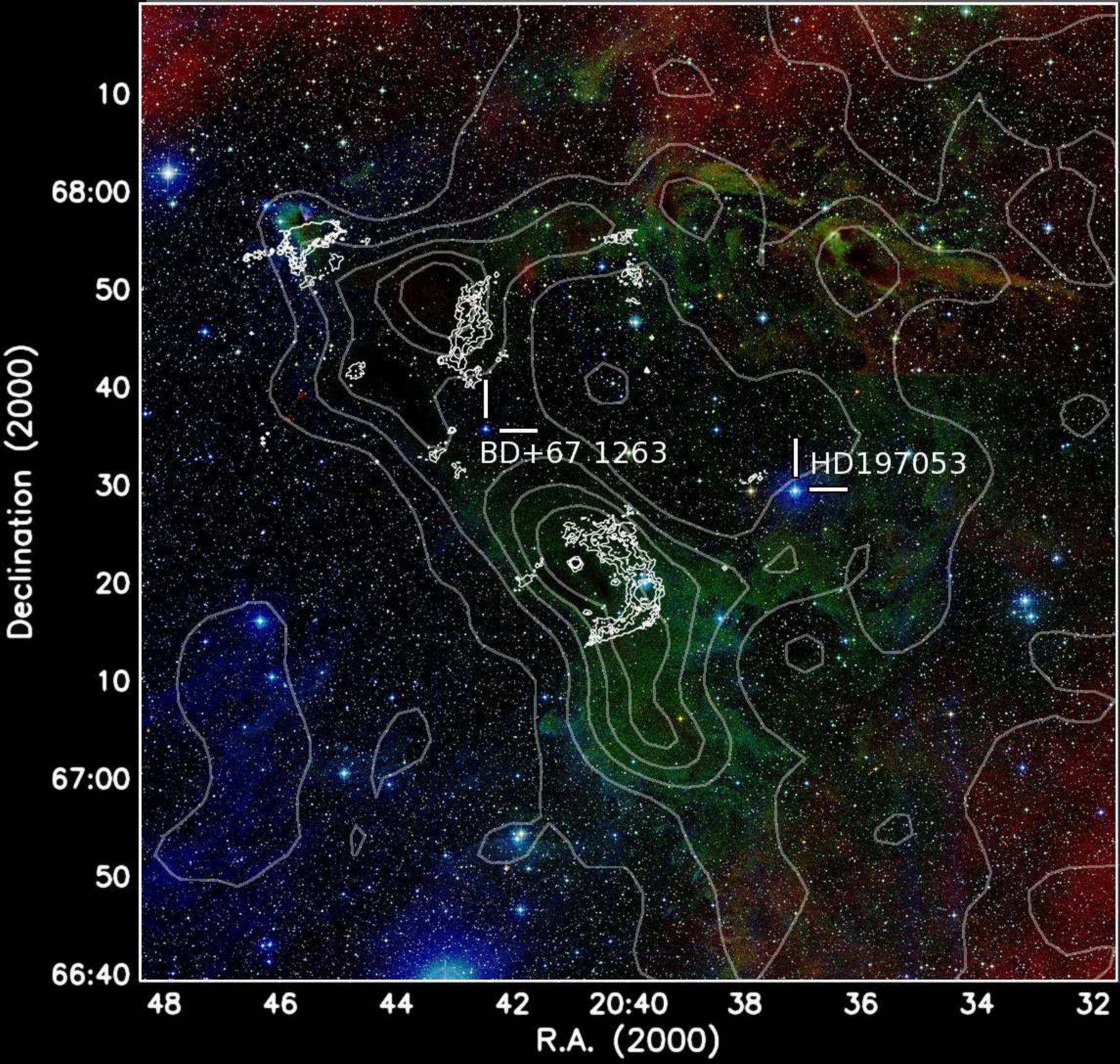}
\caption{A close-up of the extinction map (thick contours)
of the Cepheus cloud showing
the region in the box indicated in Figure~1
(from \citealt{dobashi2005}). This is the L1147-L1157 region. The
underlying image is from the Digitized Sky Survey. The
Akari 90-micron emission is superposed as thin contours 
(\citealt{nutter2009}).}
\end{figure}
%%%%%%%%%%%%%%%%%%%%%%%%%%%%%%%%%%%%%%%%%%%%%%%%%%%%%%%%%%%%%%%%%%%%%%%%%%%%%%%

We therefore hypothesise that the 90-micron emission is tracing slightly
warmer dust around the edges of these cores. This is what would be predicted 
if the cores were being externally heated. Two potential heating sources
were located. However, due to the geometry and nature of the stars involved,
we deduced that the core heating was due to the star HD197053 (\citealt{nutter2009}).

\section{Ophiuchus}

Figure 3 shows the Ophiuchus molecular cloud as seen by Herschel
at 70 to 250 microns (\citealt{pattle2015}). 
North is at the top, east is to the left.
The Oph A core can be seen clearly
at the top centre of the image, with the other cores to the south-east.
Combining these data with Herschel data allows us to calculate the
temperature of the emitting dust, and hence its mass.
We can then calculate the total mass of each core using canonical gas-to-dust
mass ratios.
Other studies have looked at the dense gas in Ophiuchus using different
tracers. For example, \citet{andre2007} studied $N2H+$ in the region.
We have used their data to calculate a virial mass for each of our cores.

%%% FIGURE %%%%%%%%%%%%%%%%%%%%%%%%%%%%%%%%%%%%%%%%%%%%%%%%%%%%%%%%%%%%%%%%%%%%
\begin{figure}
\centering
\includegraphics[width=60mm]{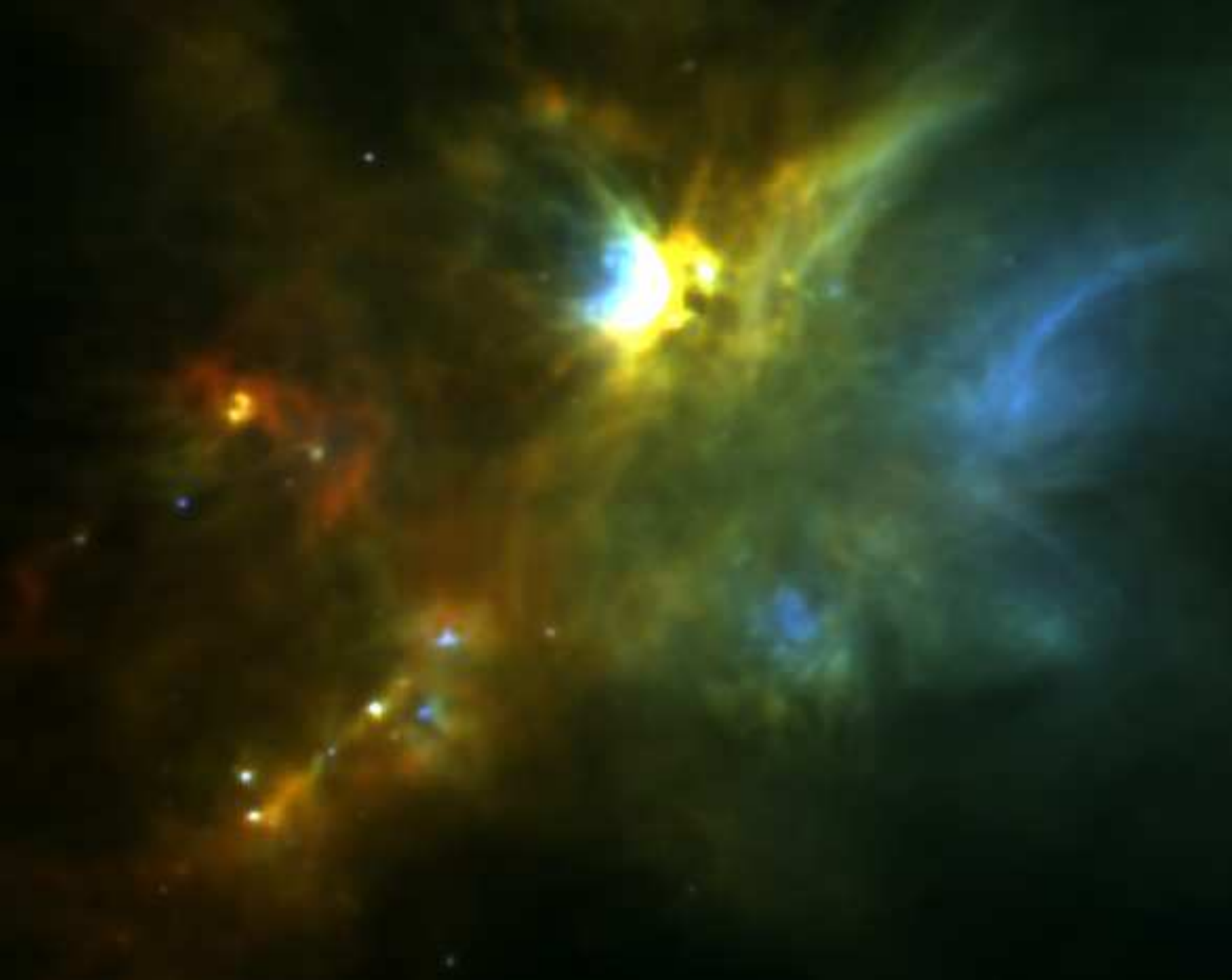}
\caption{Composite 70/160/250 image of the Ophiuchus L1688 cloud (\citealt{ladjelate2014};
 \citealt{pattle2015}). Oph A is the bright region in the upper
centre of the image.}
\end{figure}
%%%%%%%%%%%%%%%%%%%%%%%%%%%%%%%%%%%%%%%%%%%%%%%%%%%%%%%%%%%%%%%%%%%%%%%%%%%%%%%

%%% FIGURE %%%%%%%%%%%%%%%%%%%%%%%%%%%%%%%%%%%%%%%%%%%%%%%%%%%%%%%%%%%%%%%%%%%%
\begin{figure}
\centering
\includegraphics[width=60mm]{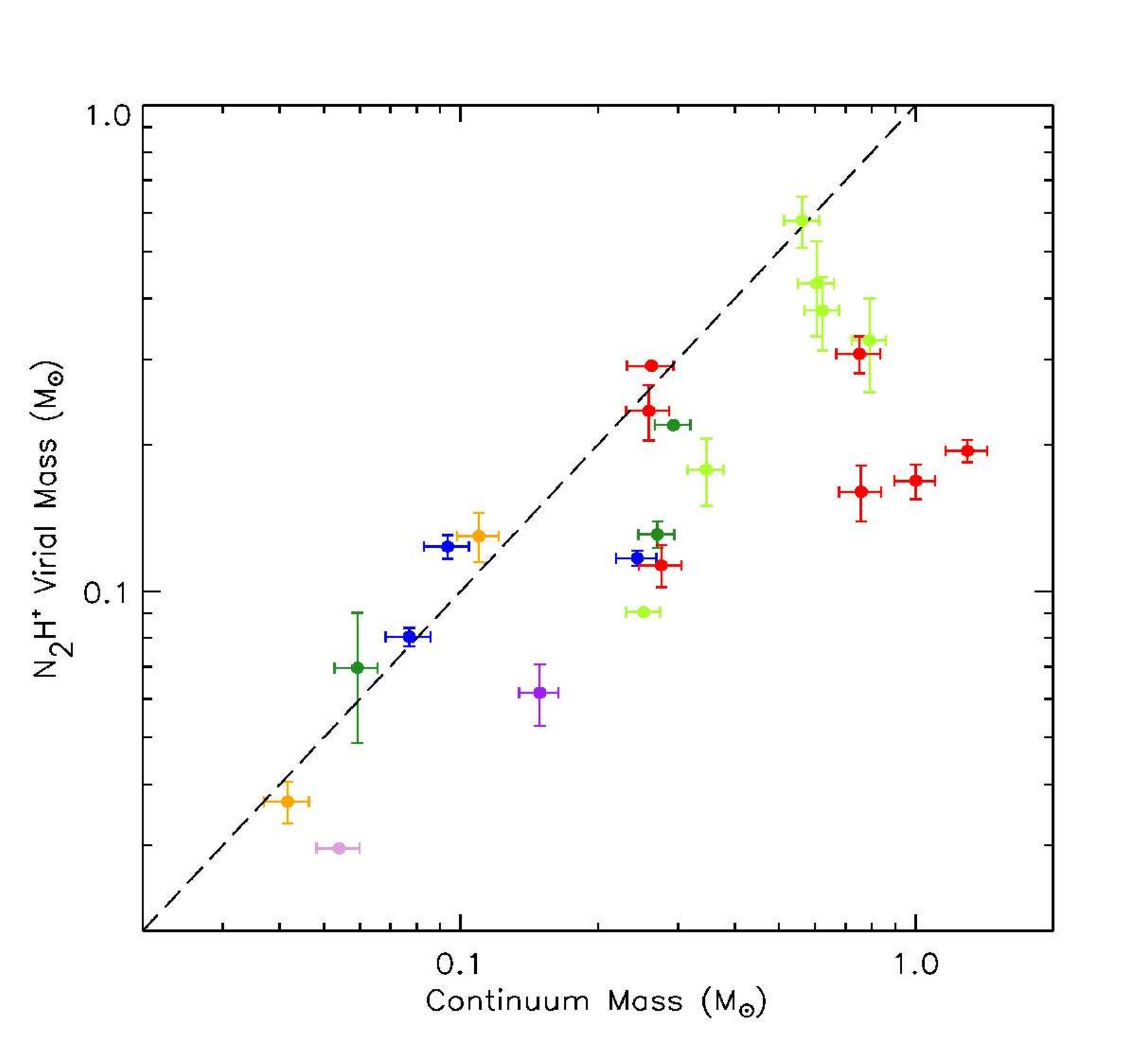}
\caption{Plot of virial mass derived from $N2H+$ against core mass derived 
from the dust emission, as measured from SCUBA2 and Herschel data 
(\citealt{pattle2015}).}
\end{figure}
%%%%%%%%%%%%%%%%%%%%%%%%%%%%%%%%%%%%%%%%%%%%%%%%%%%%%%%%%%%%%%%%%%%%%%%%%%%%%%%

%%% FIGURE %%%%%%%%%%%%%%%%%%%%%%%%%%%%%%%%%%%%%%%%%%%%%%%%%%%%%%%%%%%%%%%%%%%%
\begin{figure}
\centering
\includegraphics[width=60mm]{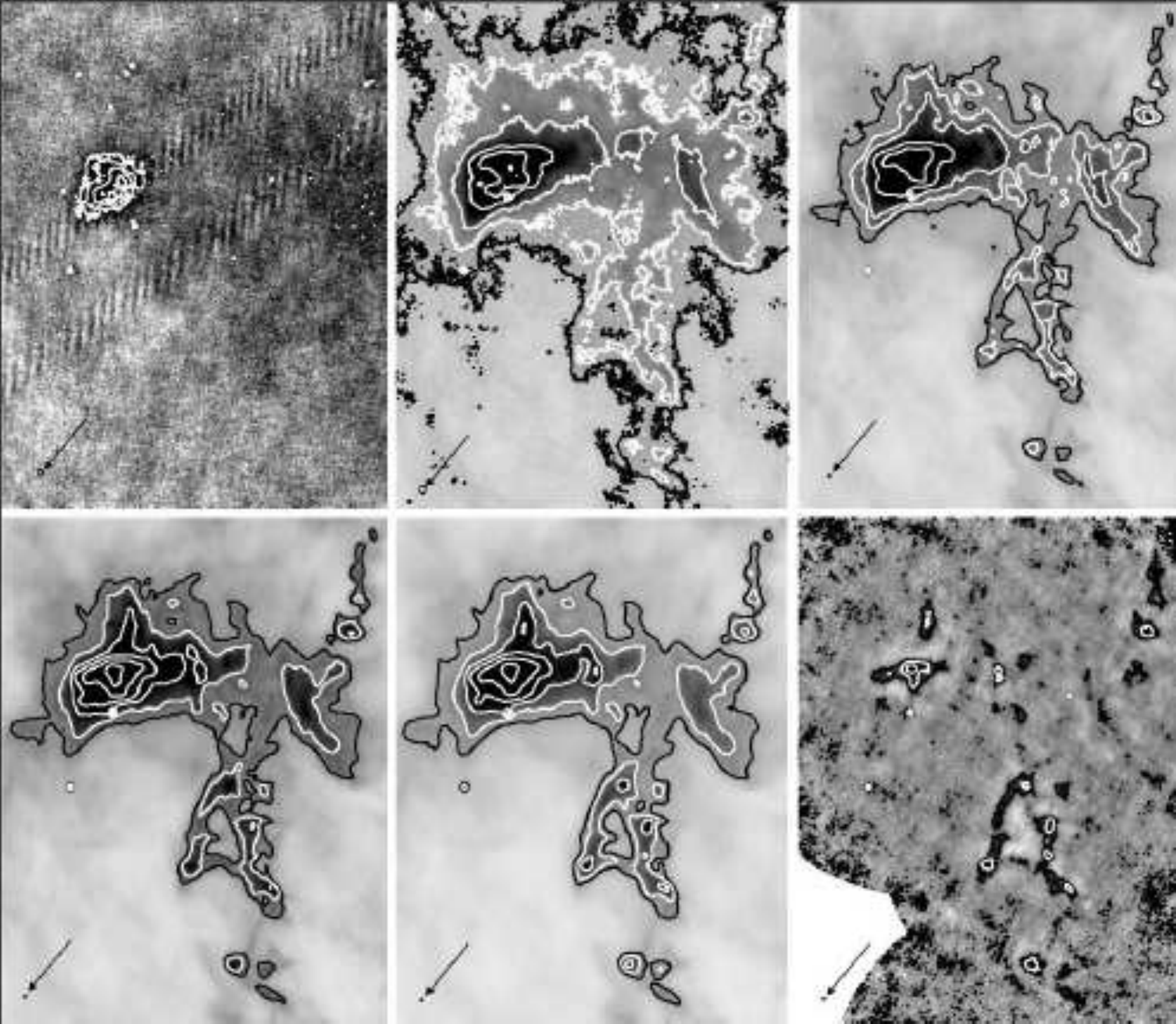}
\caption{Herschel and SCUBA2 images of the L1495 molecular cloud in Taurus.
Top row, left to right: 60, 170 and 250 microns respectively.
Lower row, left to right: 350, 500 and 850 (from SCUBA2) microns respectively.
Note how SCUBA2 only detects some of the cores seen at the other wavelengths
by Herschel (\citealt{wardthompson2014}).}
\end{figure}
%%%%%%%%%%%%%%%%%%%%%%%%%%%%%%%%%%%%%%%%%%%%%%%%%%%%%%%%%%%%%%%%%%%%%%%%%%%%%%%

%%% FIGURE %%%%%%%%%%%%%%%%%%%%%%%%%%%%%%%%%%%%%%%%%%%%%%%%%%%%%%%%%%%%%%%%%%%%
\begin{figure}
\centering
\includegraphics[width=60mm]{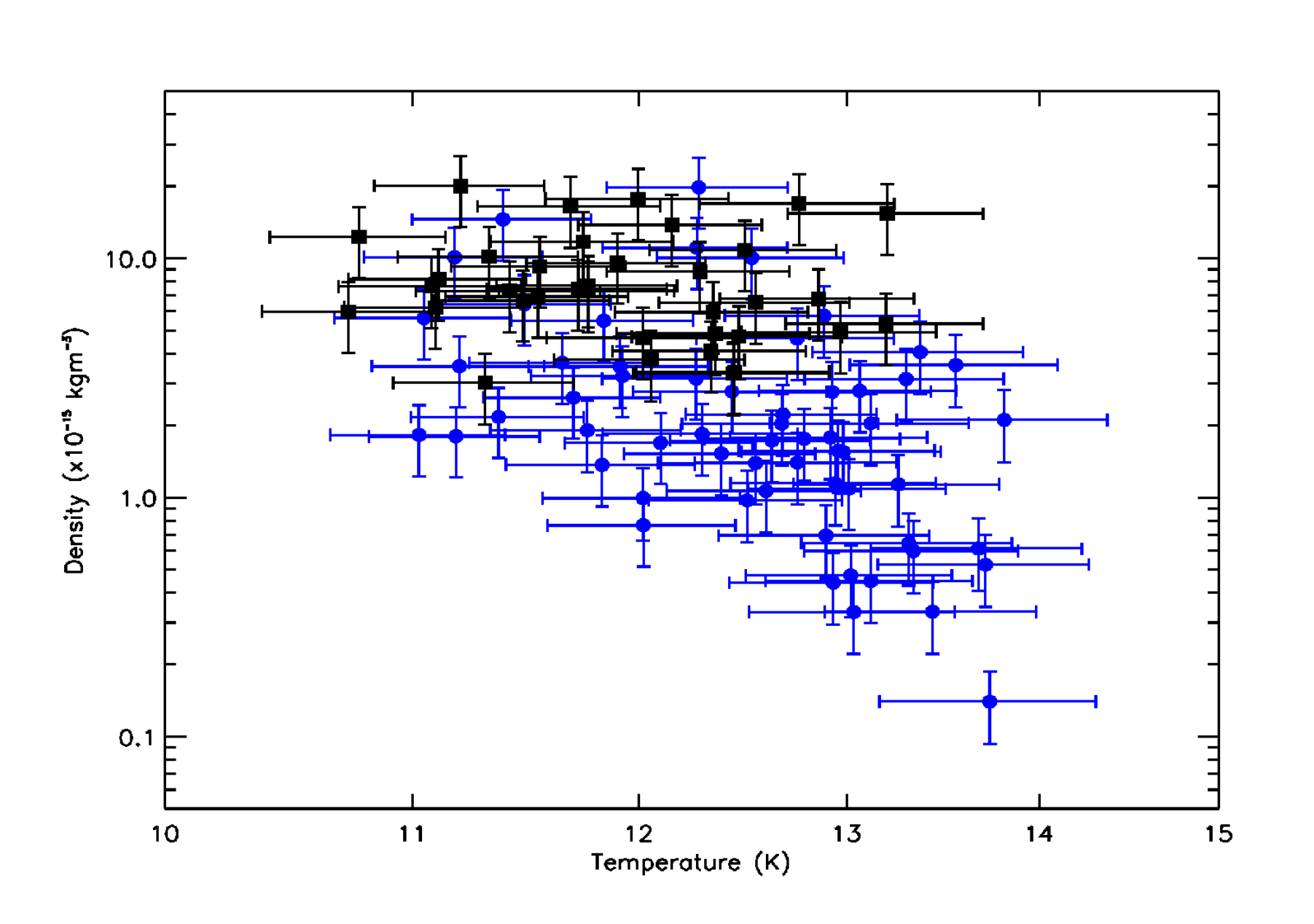}
\caption{Plot of density versus temperature for the starless cores in L1495.
Those detected by SCUBA2 (in black) all lie to the upper left-hand side of
the plot (\citealt{wardthompson2014}), indicating that SCUBA2 is only
sensitive to the coldest, densest cores.}
\end{figure}
%%%%%%%%%%%%%%%%%%%%%%%%%%%%%%%%%%%%%%%%%%%%%%%%%%%%%%%%%%%%%%%%%%%%%%%%%%%%%%%

%%% FIGURE %%%%%%%%%%%%%%%%%%%%%%%%%%%%%%%%%%%%%%%%%%%%%%%%%%%%%%%%%%%%%%%%%%%%
\begin{figure}
\centering
\includegraphics[width=60mm]{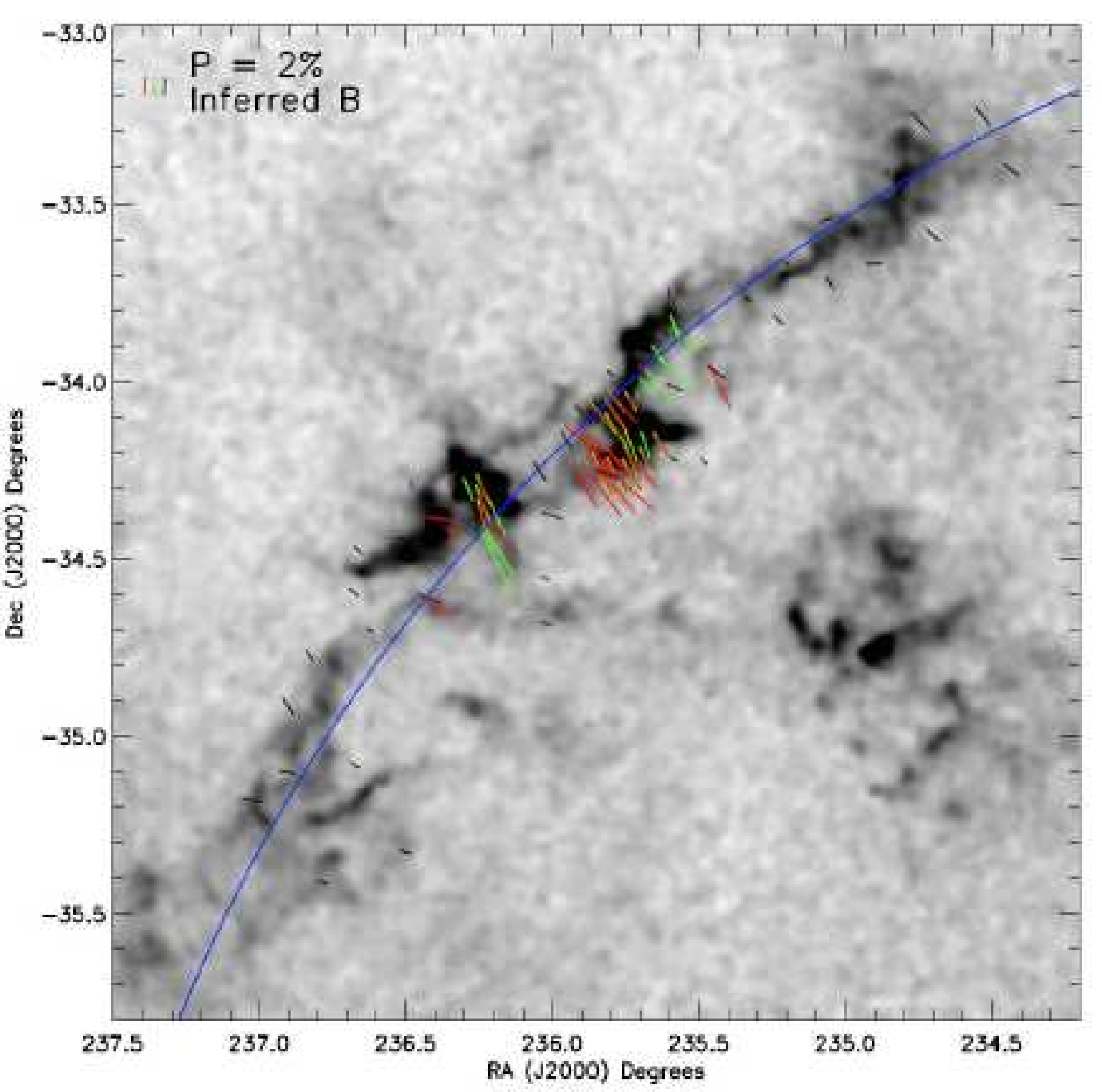}
\caption{Greyscale Herschel image of part of the Lupus I molecular cloud.
Magnetic field vectors deduced from BLAST polarisation mapping are
overlaid. The filament is traced with a curved solid line.
Note how the magnetic field lies mostly perpendicular to the direction of
the filament (\citealt{matthews2014}).}
\end{figure}
%%%%%%%%%%%%%%%%%%%%%%%%%%%%%%%%%%%%%%%%%%%%%%%%%%%%%%%%%%%%%%%%%%%%%%%%%%%%%%%

%%% FIGURE %%%%%%%%%%%%%%%%%%%%%%%%%%%%%%%%%%%%%%%%%%%%%%%%%%%%%%%%%%%%%%%%%%%%
\begin{figure}
\centering
\includegraphics[width=60mm]{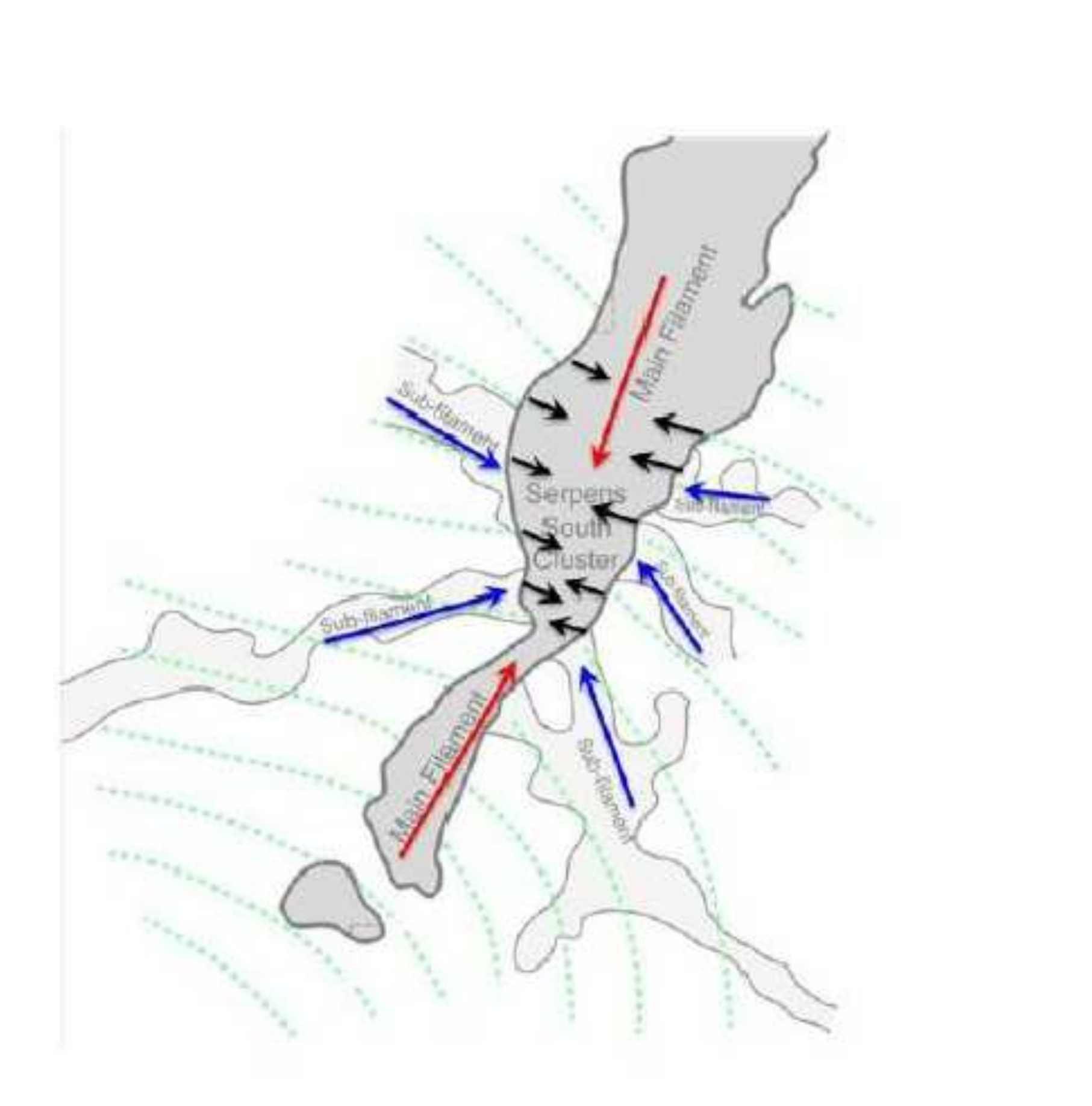}
\caption{Model of star formation consistent with the data (from \citealt{andre2014}).
 Matter flows first along magnetic field lines onto filaments. Parts
of each filament become `seed' cores. Matter then flows along the filament
onto the cores. When the mass per unit length of thefilament exceeds the
critical value the cores collapse (from \citealt{andre2014}).}
\end{figure}
%%%%%%%%%%%%%%%%%%%%%%%%%%%%%%%%%%%%%%%%%%%%%%%%%%%%%%%%%%%%%%%%%%%%%%%%%%%%%%%

%%% FIGURE %%%%%%%%%%%%%%%%%%%%%%%%%%%%%%%%%%%%%%%%%%%%%%%%%%%%%%%%%%%%%%%%%%%%
\begin{figure}
\centering
\includegraphics[width=60mm]{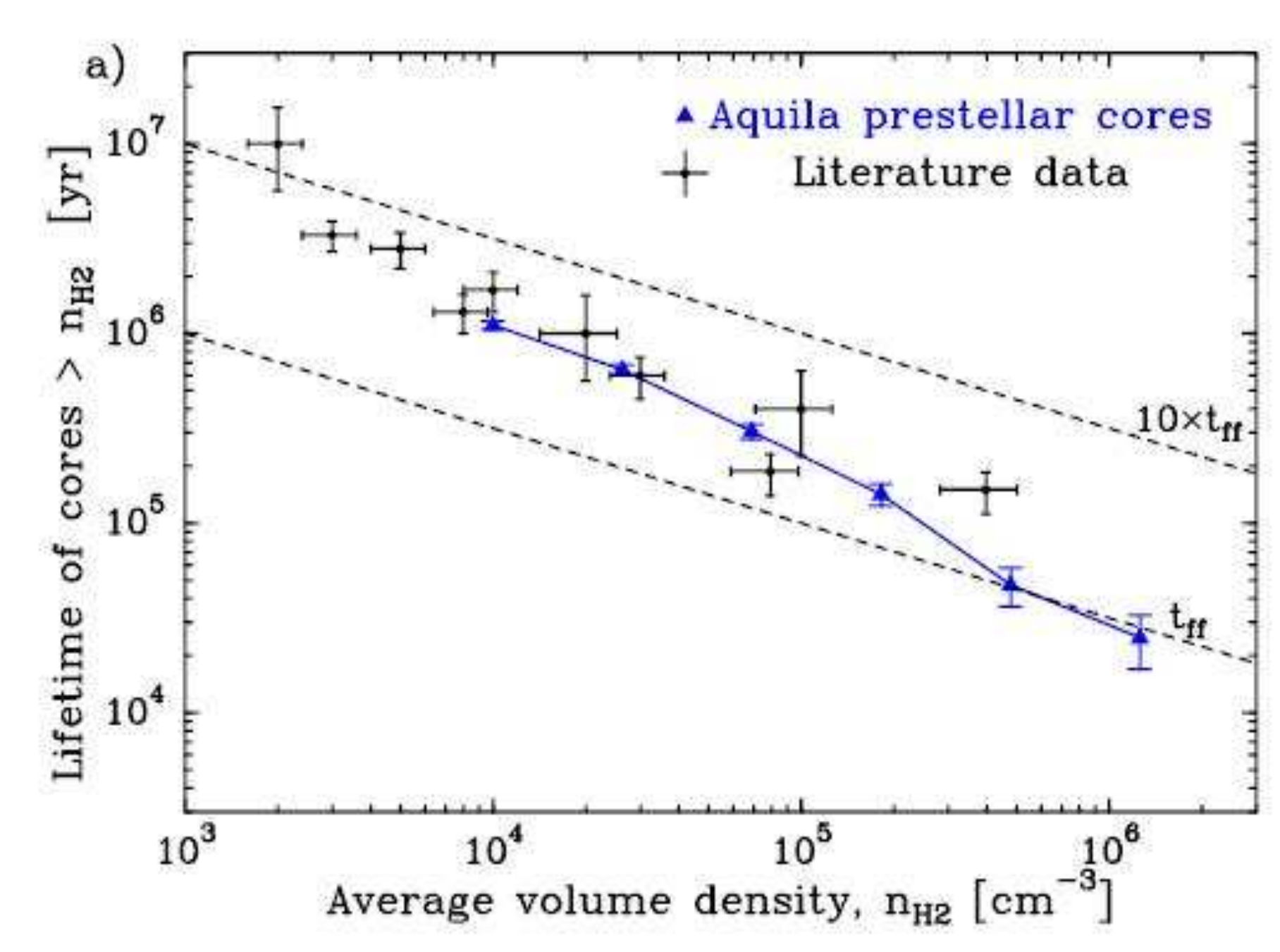}
\caption{Plot of mean lifetime versus mean column density for a sample of 
cores from the literature (\citealt{jessop2000}) and from Herschel
data on the Aquila region (\citealt{konyves2014}). The best fit to the data
is steepr than predicted by free-fall collapse (from \citealt{andre2014}).}
\end{figure}
%%%%%%%%%%%%%%%%%%%%%%%%%%%%%%%%%%%%%%%%%%%%%%%%%%%%%%%%%%%%%%%%%%%%%%%%%%%%%%%

%%% FIGURE %%%%%%%%%%%%%%%%%%%%%%%%%%%%%%%%%%%%%%%%%%%%%%%%%%%%%%%%%%%%%%%%%%%%
\begin{figure}
\centering
\includegraphics[width=60mm]{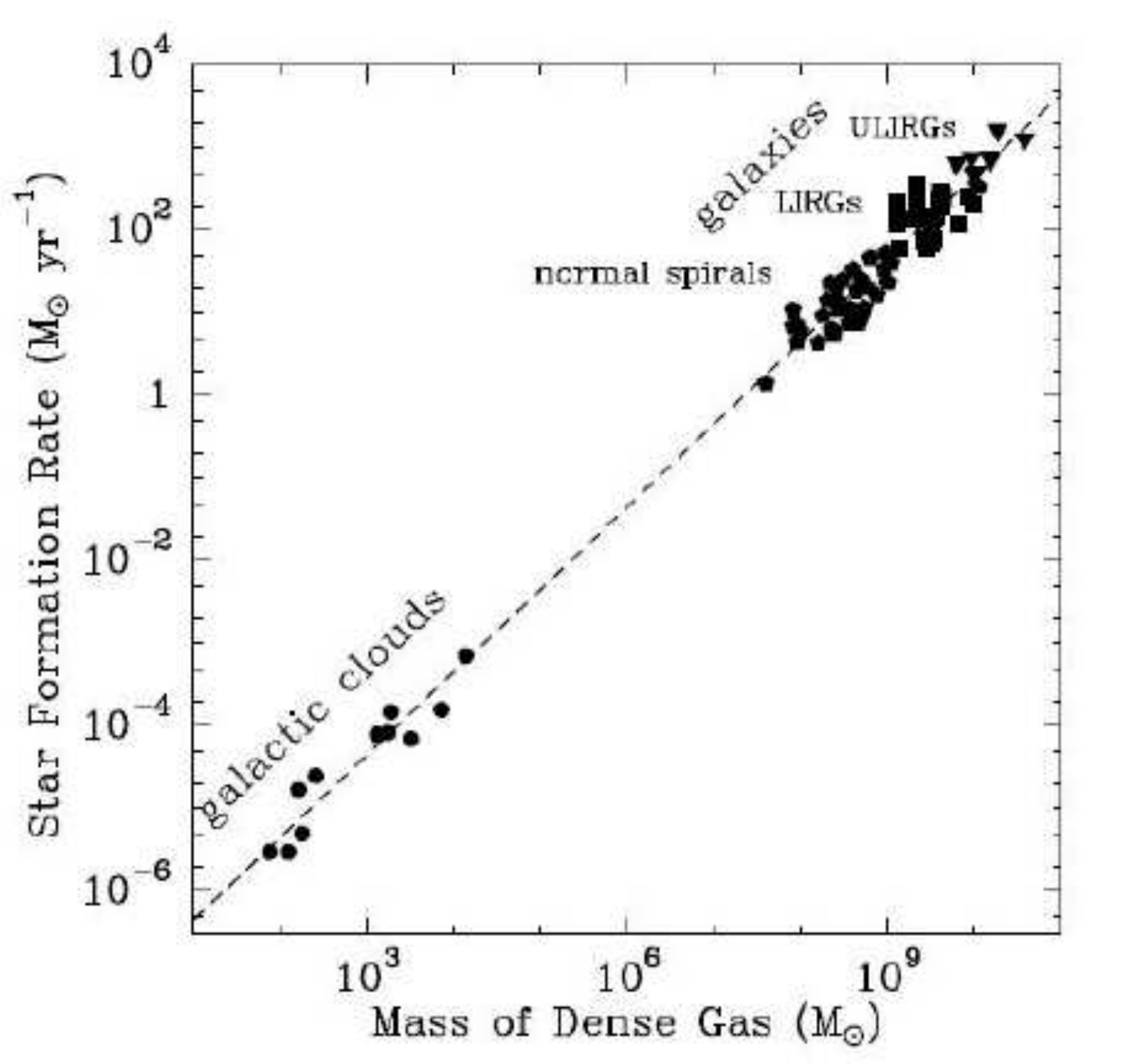}
\caption{Plot showing the relation between star formation rate and mass of 
dense gas (after \citealt{lada2012}). The dashed line is found by inverting
the best-fit line from Figure~9 (see \citealt{konyves2014}). The line is 
seen to be consistent with the data, and may provide an explanation for
the Schmidt-Kennicutt Law (from \citealt{andre2014}).}
\end{figure}
%%%%%%%%%%%%%%%%%%%%%%%%%%%%%%%%%%%%%%%%%%%%%%%%%%%%%%%%%%%%%%%%%%%%%%%%%%%%%%%

Figure 4 shows a plot of core virial mass against observed mass derived from
SCUBA2 and Herschel data (\citealt{pattle2015}). By looking at where each
individual core lies on this plot, we are able to plot a trend in the data
from upper right to lower left in this plot, which correlates with a trend in
location from north-west to south-east in Figure~3.

We can explain this in terms of the degree of gravitational boundedness of
the cores. Those to the western side of Figure~3 are more tightly 
gravitationally bound than those to east. We interpret this in terms of 
sequential star formation from west to east in this image (\citealt{pattle2015}). 

The more tightly gravitationally bound cores to the west are 
interpreted as being more evolved than those towards the east. This
is consistent with a previously proposed hypothesis for the star formation
in Ophiuchus, comparing L1688 to L1689 (\citealt{nutter2006}).

These latter
authors found that star formation in L1688 had progressed much further
than in L1689, referring to L1689 as the `dog that didn't bark'.
This was interpreted as being caused by the external influence of the 
Upper Sco OB
association to the west of Ophiuchus triggering star formation in L1688
before L1689. This trend is exactly consistent with what we see in the
SCUBA2 and Herschel data (\citealt{pattle2015}).

Therefore, in both Cepheus and Ophiuchus we see the evolution of cores 
being heavily influenced by their surroundings and their environment.
In particular, in Ophiuchus we are seeing an apparent case of sequential
star formation from west to east across the region.

\section{Taurus}

Figure 5 shows the L1495 region in Taurus
(\citealt{wardthompson2014}). Five Herschel wavebands and one waveband
from SCUBA2 are shown. A number of cores and filamentary structures can
be seen. In particular a triangle-shaped arrangement of filaments can be 
seen just below centre of each image. The ubiquity of filaments is a
feature seen in almost all of the Herschel data of star-forming regions
(\citealt{andre2014}).

The bright region towards the upper part of the image
that is seen at most wavebands is the L1495A core, and is exactly
coincident with L1495A-S. 
This lies at the head of a large, filamentary structure that is seen
clearly in the Herschel data of the Taurus region (\citealt{marsh2014}).
We measured a temperature 
gradient across this core, with the hotter material lying to the south
(\citealt{wardthompson2014}).

There is a bright star, slightly to the south of L1495A-S, which is 
known as V892 Tau (IRAS04155+2812). This is a Herbig Ae/Be star, and 
it is clearly heating L1495A-S, which is otherwise starless, and causing
the temperature gradient across the core. This is very similar to that seen 
in the Cepheus discussed above. So once again we see that environment is
affecting evolution.

Note that only some of
the cores visible in the Herschel images are seen in the SCUBA2 image.
These regions are seen most clearly by SCUBA2 at 850~$\mu$m, 
but not seen so clearly at shorter wavelengths by Herschel. Note, for example
the filament to the upper right in the Herschel images,
which appears to Herschel 
as no different from the other filaments, but which is almost
invisible in the SCUBA2 images. 

Figure 6 shows a plot of density versus temperature for the cores in L1495.
The cores not detected by SCUBA2 are indicated by a different symbol from
those thare detected by SCUBA2. Note that
the cores not detected by SCUBA2 all lie to the lower right-hand side of
this plot (\citealt{wardthompson2014}). This indicates that SCUBA2 is only
sensitive to the coldest, densest cores, whereas Herschel sees all of the
cores.

\section{Lupus}

Figure 7 shows a Herschel image of the Lupus I molecular cloud.
Magnetic field vectors deduced from BLAST polarisation mapping are
overlaid (\citealt{matthews2014}). 
The filament is traced with a curved solid line.
Note how the magnetic field lies mostly perpendicular to the direction of
the filament (\citealt{matthews2014}).

This is consistent with other observations of filaments and magnetic fields
(\citealt{palmerim2013}). This has led to a model being proposed for star
formation that invokes the magnetic field funnelling material onto filaments
(\citealt{andre2014}).
Material then flows along filaments to form cores
(\citealt{balsara2001}; \citealt{andre2014}). When sufficient mass has
accreted in a core that the filament's critical mass-to-flux ratio is exceeded
the core collapses (\citealt{inutsuka1997}).

The observation that the majority of cores form on filaments
(\citealt{konyves2010}) indicates that this is the dominant
mode of star formation. Observation of flows along filaments has been
observed before (e.g. \citealt{balsara2001}), but the new observations
give the first indication of flow onto filaments (\citealt{palmeirim2013}).

\section{Discussion}

Figure 9 shows a plot of lifetime versus mean density for samples of dense
cores that was originally proposed by \citet{jessop2000}.
A general trend was observed that was steeper than that predicted by free-fall
collapse. More recent Herschel data also show this same trend 
(\citealt{konyves2014}).

Andr\'e et al. (2014) have proposed that this trend may in fact explain 
the Schmidt-Kennicutt Law, in its recent incarnation by \citet{lada2012}.
If the best-fit line from Figure~9 is converted into a star formation rate
and account is taken of the filamentary nature of the clouds
of dense gas above the 
star-forming threshold (\citealt{lada2012}), then for Aquila we 
derived a star formation rate, given by 
$SFR = 4.5 \times 10^{-8} M_\odot yr^{-1} \times
(M_{\rm dense}/M_\odot)$ (\citealt{konyves2014}).

Figure 10 shows a plot of star formation rate against mass of dense gas
above the star-forming threshold (after \citealt{lada2012}), showing both
Galactic clouds and external galaxies from normal spirals to ULIRGs
(\citealt{gao2004}).
Also shown on this plot as a dashed line is the star formation mentioned 
above, as derived from Herschel data of Aquila (\citealt{konyves2014})
and earlier data (\citealt{jessop2000}).

The dashed line fits exactly to the data of both our own Galaxy and of 
external galaxies. This led us to hypothesise that the filamentary nature
of molecular clouds, and the core life-time versus density relation, may
ultimately lead to a universal law for star formation (\citealt{andre2014}).

\section{Conclusions}

We have shown data from Akari, Herschel and SCUBA-2 of pre-stellar cores 
in the star-forming regions in Cepheus, Ophiuchus, Taurus and Lupus. A
number of themes have emerged:

\begin{itemize}

\item We have seen that the environment in which the cores exist 
is extremely important. For example, in Cepheus and Ophiuchus the external 
effects of nearby luminous stars is causing temperature gradients across
the clouds. In the case of Ophiuchus we are also seeing sequential star
formation.

\item In Taurus we have compared and contrasted the Herschel data with the 
SCUBA2 data and seen that whereas Herschel is seen to be sensitive to all
structures, SCUBA2 only picks up the coldest, densest cores that are
probably pre-stellar in nature.

\item In Taurus and Lupus we have seen examples of filamentary molecular
clouds that are seen to be everywhere in the Taurus data. In fact, the
Herschel data show that core formation on filaments is the dominant mode of
star formation (\citealt{andre2010}).

\item We have seen how the magnetic field in Lupus lies perpendicular to
the main filament, supporting a model of
star formation in which the magnetic field funnels material onto filaments,
and material flows along the filaments onto cores (\citealt{andre2014}).

\item We have seen how observations of the life-time versus density
relation, together with Herschel observations of Aquila, lead to a new
explanation of the star-formation versus dense gas mass relation
(\citealt{andre2014}).

\end{itemize}

It is clear that these far-infrared and submm instruments have 
changed our views about star formation.

%%% ACKNOWLEDGMENTS (IF ANY) %%%%%%%%%%%%%%%%%%%%%%%%%%%%%%%%%%%%%%%%

%\acknowledgments

%We are grateful to ....

%%% APPENDICES (IF ANY) %%%%%%%%%%%%%%%%%%%%%%%%%%%%%%%%%%%%%%%%%%%%%

%\appendix
%\section{Appendix Title}

%Some text.

%%% CALL LIST OF REFERENCES (natbib STYLE) %%%%%%%%%%%%%%%%%%%%%%%%%%

\end{document}